\documentclass[showpacs,aps,pra,twocolumn,amsmath,amssymb]{revtex4}

\usepackage{graphicx}
\usepackage{color} 
\usepackage{cancel}
\usepackage{ulem}
\usepackage{soul}
\definecolor{orange}{rgb}{1,0.5,0}

\begin{document}

\title{Entanglement dynamics in presence of diversity under decohering environments}

\author{Fernando Galve}
\author{Gian Luca Giorgi}
\author{Roberta Zambrini}
\affiliation{IFISC (UIB-CSIC),
Instituto de F\'isica Interdisciplinar y Sistemas Complejos, UIB Campus,
E-07122 Palma de Mallorca, Spain}

\date{\today}

\begin{abstract}

We study the evolution of entanglement of a pair of coupled, non-resonant harmonic oscillators in contact with an
environment. For both the cases of a common bath and of two separate baths for each of the oscillators, a full
master equation is provided without rotating wave approximation. This allows us to characterize the entanglement
dynamics as a function of the diversity between the oscillators frequencies and their mutual coupling.  Also the
correlation  between  the occupation numbers is considered  to explore the degree of quantumness of the system. 
The singular effect of the resonance condition (identical oscillators) and its relationship with the possibility of preserving asymptotic
entanglement are discussed. The importance of the bath's memory properties is investigated by comparing Markovian
and non-Markovian evolutions. 
\end{abstract}


\maketitle

\section{Introduction}

Coupled harmonic oscillators are the first approximation to a broad class of extended
systems not only in different fields of physics but also in chemistry and biology. Within
the quantum  formalism, they are the basis of the description of electromagnetic field
interactions in quantum optics and approximate lattice systems in different
traps in atomic physics \cite{loudon,haroche}. Moreover, in the last few years there has
been impressive progress towards  the cooling and back-action evasion measurement of
`macroscopic' -in terms of number of atoms- harmonic oscillators, allowing the observation
of their quantum behavior. Two main class of systems experimentally realized are 
nanoelectromechanical structures (NEMS) \cite{Naik,Schwab} and different kinds of optomechanical
systems  \cite{optomech} where nano- and micromechanical devices,   cavities or  suspended
mirrors are respectively coupled  to single electrons or light. As an example, it has recently 
been reported the observation of NEMS extremely near to the ground
state of motion with an occupation factor of just 3.8 \cite{Schwab}. These experiments
would allow to observe coherent quantum superposition states, entanglement and to study
decoherence processes in a controllable way on massive solid-state objects.

Phenomena associated to the $coupling$ of these quantum oscillators have been revisited
within the context of quantum information in many theoretical studies during the last
decade \cite{entinarrays,harm_chains,suddenstuff,galve,liu,paz-roncaglia,paz-roncaglia2}.
Entropy and entanglement in extended systems with many degrees of freedom (harmonic
chains or lattices)\cite{entinarrays}, have been characterized in fundamental and thermal
states, exploring scaling laws and connections with phase transitions
\cite{harm_chains}.  An important advantage is that  these systems admit Gaussian state
solutions having a well defined \cite{simon} computable \cite{vidal} measure of
entanglement, the logarithmic  negativity \cite{volume}. The question of the generation
of entanglement has also been addressed considering oscillators whose  parameters are
modulated in  time  \cite{suddenstuff,galve}. In these studies losses were generally
neglected, while recently the effects of decoherence on a pair of entangled oscillators
have been considered in presence of dissipation through baths of infinite oscillators
\cite{liu,paz-roncaglia,paz-roncaglia2}. Our aim in this paper is to analyze a rather
unexplored aspect of this problem that is {\it the effect of the diversity} on the
entanglement between coupled harmonic oscillators in different situations. Indeed,
instead of considering identical oscillators, we look at the effects of detuning between
their frequencies, $\omega_1$ and $\omega_2$. 

The interest about diversity effects on entanglement is both theoretical and related to
experimental issues. We mention, for instance, the effect of diversity of frequencies of
two photons entering in a beam splitter. This leads to a completely different output with
respect to the case of indistinguishable photons \cite{Hong1987,Yamamoto} 
\footnote{ Notice also that recently the robustness of equal light modes quantum interference has
been tested against another kind of diversity -on states-, being dissimilar source considered
\cite{Bennett2009}.} 
Coupling between different harmonic modes has been also
extensively studied in  quantum optics in presence of nonlinear interactions and
parametric coupling, allowing, for instance, generation of entanglement between photons
pairs and intense light beams \cite{squeezingbook,braunstein} of different colors. In
that context, however,  the frequency diversity of each pair of oscillators is
compensated by a third mode and is in general not relevant, while here we focus on
pairs of mechanical oscillators off-resonance and with constant couplings. The main
expected consequence is indeed an effective decoupling of the oscillators due to fast
rotation of their interaction term and we will show the effects on the robustness of
entanglement. 

It is clear that the identity of the oscillators in general is a very peculiar and strong
assumption not always justified and introducing a symmetry into the system with deep
consequences. It is indeed important to clarify the effects of relaxation of this
symmetry introducing some diversity. In an extension from two coupled oscillators to an
array, this will imply a break in the translational symmetry. Apart from the fundamental
interest, we point out that in experimental realization of engineered arrays of massive
quantum oscillators, some diversity between them might actually be  unavoidable. Coupled
oscillators with different frequencies have been also suggested for quantum limited
measurements \cite{Leaci}. Finally, in many experiments coupled oscillators actually
model different physical entities (for instance radiation and a moving mirror in
optomechanics) and a symmetric Hamiltonian would describe only a very special case
\cite{aspelmayer}. 

Our analysis of entanglement evolution encompasses both diversity between harmonic
oscillators and dissipation. Once the oscillators are prepared in some entangled state
(for instance through sudden switch of their coupling), we look at its robustness increasing
diversity and coupling strength. It is well known that an object  whose all degrees of
freedom are coupled to an environment will  decohere  into a thermal state with the same
temperature as the heat bath \cite{haroche}. The thermal state, unless temperatures are very
low,  is separable and highly entropic, so that after thermalization all entanglement shared
between the coupled oscillators disappears \cite{anders}. As the dissipating  oscillators
reach a separable state in a finite time, and not asymptotically, the name of sudden death
has been suggested \cite{suddendeath}. Still under certain conditions on the ways in which
the oscillators dissipate, entanglement can survive asymptotically \cite{asympt_ent_osc}. 
The transition from a quantum to a classical behavior in these systems does not convey only
fundamental interest being also important in view of applications of harmonic systems
operating at the quantum limit, including  quantum information processing and not shot-noise
limited measurements of displacement, forces or charges \cite{applications}. These phenomena
have generally been studied for identical oscillators and, in particular, previous works
provide a master equation description in the case of a fully symmetric Hamiltonian
\cite{liu,paz-roncaglia}. Off-resonance oscillators in presence of a common bath have
recently been considered in Ref. \cite{paz-roncaglia2}  showing that in this case lower
temperatures are needed to maintain  entanglement asymptotically. In this work we analyze
systematically the role of diversity on the entanglement dynamics and provide the full
master equations for $\omega_1\neq\omega_2$ (i) both for common and separate baths for the
two oscillators, (ii) without the approximation of rotating waves, generally assumed when
modeling light fields, and (iii) comparing results to the non-Markovian case. 

Recently there have been some works focusing on memory effects and  non-Markovianity. In
the case of continuous variables, non-Markovian effects on entanglement evolution  have
been discussed \cite{maniscalco,liu}. In Ref. \cite{maniscalco}, the authors  considered
two identical oscillators coupled to separate baths in the very high temperature regime
and analyze how the matching between the frequency of the oscillators and the spectral
density of the bath affects both Markovian and non-Markovian dynamics. In Ref. \cite{liu},
the non-Markovian evolution has been compared with the case where both the Markovian limit
and the rotating wave approximation in the system-bath coupling are taken. Since the
effects of these two approximations are not easily separable, it is difficult to single
out non-Markovian corrections from that analysis.  Our study of entanglement dynamics
in presence of diversity is also extended, for the sake of comparison, to the
non-Markovian case, looking at both common and separate baths and showing that deviations
are actually negligible.

In Sect. \ref{model} we introduce the model of a pair of oscillators with different
frequencies and coupled through position both between them and with the baths of
oscillators. Both master equations for one common and two separate baths are presented,
providing all the details in the Appendix. Temporal decay of entanglement between the
oscillators and correlations between the occupation numbers are shown  in Sect.
\ref{results} analysing the role of frequency diversity when also their coupling
strength is varied. Apart from the entanglement dynamics, we also discuss its robustness 
(asymptotic entanglement for common bath) in the context of the symmetry of the system,
Sect. \ref{asympt}.  Non-Markovian deviations from these results are shown in Sect.
\ref{non-Markovian} and further discussion and conclusions are left for Sect. \ref{concl}.

\section{Model and master equations}
\label{model}

We consider two harmonic oscillators with the same mass and different frequencies
coupled to a thermal bath. As discussed in Ref.\cite{zell}, depending on the
distance between the two oscillators, different modelizations of the system-bath
interaction can be done. We will discuss the case of two distant objects, which
amounts to considering the coupling with two independent baths, and compare it with
the zero-distance scenario (common bath).  Analyzing the role of
diversity on the quantum features of this system, we also generalize previous
works on identical oscillators dissipating in common and separate baths
\cite{liu,paz-roncaglia,paz-roncaglia2}.
 
\subsection{Separate baths}

The model Hamiltonian, with each oscillator coupled to an infinite number of
oscillators (separate baths), is $H^{sep}=H_S+H_B^{sep}+H_{SB}^{sep}$.
The system Hamiltonian
\begin{equation}
H_{S}=  \frac{p_1^2}{2}+\frac{1}{2}\omega_{1}^2x_1^2+ \frac{p_2^2}{2}+
\frac{1}{2}\omega_{2}^2x_2^2 +\lambda x_1 x_2
\end{equation}
describes two oscillators with different frequencies $\omega_{1,2}$, 
coupled through their positions,  
\begin{equation}
 H_{B}^{sep}=\sum_{k}\sum_{i=1}^2\left(\frac{P_k^{(i)2}}{2}+\frac{1}{2}  
 \Omega_{k}^{(i)2}  X_{k}^{(i)2}\right)          
\end{equation}
 is the free Hamiltonian of two (identical) bosonic baths, and 
\begin{equation}
H_{SB}^{sep}=\sum_{k}\lambda_{k}^{(1)} X_{k}^{(1)}x_1+\sum_{k}\lambda_{k}^{(2)}
X_{k}^{(2)}x_2\label{sb}
\end{equation}
encompasses the system-bath interaction. 

The master equation for the reduced density matrix of the two oscillators, up to the
second order in $H_{SB}^{sep}$ (weak coupling limit) is, assuming $\hbar=1$,
\begin{eqnarray}
 \frac{d \rho}{dt}&=&-i[H_S,{\rho}]\nonumber\\&-& 
 \frac{1}{2}\sum_{i,j=1}^2 \big\{i\epsilon_{ij}^2[x_i x_j,\rho]+D_{ij}[x_i, [x_j,\rho]]
 \nonumber\\ &+&
 i\Gamma_{ij}[x_i,\{ p_j,\rho\}]-F_{ij}[x_i ,[p_j,\rho]]\big\}.\label{me_sep}
\end{eqnarray}
In Appendix we give an explicit derivation of the master equation together with the definition
of the coefficients. Then, $\rho$ is subject to energy renormalization ($\epsilon_{ij}$),
dissipation $\Gamma_{ij}$, and diffusion $D_{ij},F_{ij}$. 
Coefficients in Eq. (\ref{me_sep}) depend  on time, but in the following, we
will consider the Markovian limit (performed sending $t\rightarrow \infty$ in Eqs.
(\ref{coe}-\ref{cog})). Non-Markovian corrections will be discussed in Sec.
\ref{non-Markovian}. We anticipate that we will focus on the Markovian limit because
corrections dropping from such an approximation  turn out to be negligible in most of the
cases, both for equal or different frequencies $\omega_1$ and $\omega_2$, and for common or
separate baths.

To obtain an explicit expression for $\epsilon_{ij},D_{ij},\Gamma_{ij},F_{ij}$,
we need to know the density of states of the baths, 
defined for both of them as
\begin{equation}\label{eq:J}
 J(\Omega)=\sum_{k}\frac{\lambda_{k}^{2}}{ \Omega_{k}}\delta(\Omega-\Omega_k).
\end{equation} 
Here, we will consider explicitly the Ohmic environment with a Lorentz-Drude
cut-off function, whose spectral density is
\begin{equation}\label{JOhm}
 J(\Omega)=\frac{2 \gamma}{\pi}\Omega\frac{\Lambda^2}{\Lambda^2+\Omega^2}
\end{equation} 
We learned from the master equation (\ref{me_sep}) that bare frequencies of the
oscillators are renormalized because of the presence of $\epsilon_{ij}$, and
this renormalization turns out to depend on the frequency cut-off $\Lambda$. This
undesirable unphysical 
 effect can be removed by adding to the initial Hamiltonian
counter-terms \cite{caldeiraleggett}
which exactly compensate the asymptotic
values of  $\epsilon_{ij}$. 
 This is accomplished by replacing
$\epsilon_{ij}(t)$ with $\epsilon_{ij}(t)-\epsilon_{ij}(\infty)$
in Eq. (\ref{me_sep}). 
Then, in the Markovian case, we will simply drop them from the master equation. This
renormalization procedure amounts to redefining the natural frequency in terms of
the observed frequency. 

\subsection{Common bath}

We introduce now the case of common bath modeled by 
\begin{equation}
 H_{B}^{c}=\sum_{k}\left(\frac{P_k^{2}}{2}+\frac{1}{2}  
 \Omega_{k}^{2}  X_{k}^{2}\right)          
\end{equation}
and we consider the same spectral density (\ref{JOhm}). The interaction term reads
\begin{equation}
H_{SB}^c=\sum_{k}\lambda_{k} X_{k}(x_1+x_2),\label{sb_common}
\end{equation}
meaning that the system is coupled to the bath through the mode  
$x_+=(x_1+x_2)/\sqrt{2},$   while
the mode  $x_-=(x_1-x_2)/\sqrt{2}$  is decoupled. The master equation in this case
is
 \begin{eqnarray}\label{me_comm}
 \frac{d \rho}{dt}&=&-i[H_S,{\rho}]- \frac{i}{2} \sum_{i\neq j}(\bar{\epsilon}_{ii}^{2}-\bar{\epsilon}_{jj}^2)x_j \rho x_i\nonumber\\
&&-\frac{1}{\sqrt{2}}\sum_{i=1}^2\big\{ i \bar{\epsilon}_{ii}^2[x_ix_+, \rho]
 +{\bar D_{ii}}[x_+,[x_i,\rho]]\nonumber\\
&&+i\bar\Gamma_{ii}[x_+,\{p_i,\rho\}]-\bar F_{ii}[x_+,[p_i,\rho]]\big\},
\end{eqnarray}
(see the Appendix for the definition of coefficients).
Coupled systems dissipating in a common bath have been recently subject of interest
because of the possibility to preserve entanglement asymptotically at high temperatures,
being this a major distinctive feature with respect to the case of separate baths.  In
Sect. \ref{asympt}, we will study entanglement evolution under Eq. (\ref{me_comm}) 
revisiting the possibility to maintain entanglement asymptotically in presence of
diversity  (frequency detuning).  As a matter of fact, whereas in the case of equal
frequencies the decoupled mode $x_-$ is  an eigenmode of the isolated system, this is no
longer true once $\omega_1 \neq \omega_2$.  We will see how asymptotic entanglement
 can be preserved in presence of diversity, through a proper choice of the
system-bath coupling constants. 

Another difference in the case of common bath is the lack of the symmetry
$\lambda\rightarrow-\lambda$.  A closer look to the form of the total Hamiltonian in
presence of separate baths allows us to single out its symmetry properties.
$H^{sep}$ is clearly modified by the transformation $\lambda\rightarrow-\lambda$. However,
once the canonical transformation $U$ is introduced,
 such that $U x_2 U^\dag=-x_2 $ and  $U p_2
U^\dag=-p_2 $, it is easy to show that the master equation is invariant under the combined
action of $U$ and the flip of $\lambda$. In fact, while $H_{S}$
is left unmodified, the change in $H_{SB}^{sep}$ (that would correspond to flip
$\lambda_{k}^{(2)}$ in  $-\lambda_{k}^{(2)}$) is ineffective with respect to the evolution
of the reduced density matrix, being $\rho$ determined only by contributions proportional to
$(\lambda_{k}^{(i)})^2$. A simple consequence of this symmetry can be observed studying
the evolution of the two-mode squeezed state 
 \begin{eqnarray}\label{TMS}
| \Psi_{TMS}
\rangle=\sqrt{1-\mu}\sum_{n=0}^\infty \mu^{n/2}|n\rangle|n\rangle, 
 \end{eqnarray}
where $\mu=\tanh^2 r$,
and $r$ is the squeezing amplitude. For this state, the canonical transformation $U$
amounts to changing $r$ in $-r$. Then, the same evolution must be obtained considering
$\lambda$ in  $H^{sep}$ and a given squeezing $r$ in the initial condition, or $-\lambda$
and $-r$. All these considerations are true only in the case of separate baths. Once the
common bath is taken, because of the action of $U$, the mode coupled to the bath is $x_-$
instead of $x_+$.  Therefore, in the case of separate baths,  results for opposite
$\lambda$ are equivalent to a change in the sign of the initial squeezing, while this is
not the case 
for a common bath.

\section{Entanglement and quantum correlations}
 \label{results}

The calculation of bipartite entanglement for mixed states is generally an unsolved task.
Nevertheless, the criterion of positivity of the partial transposed density matrix of
Gaussian two-mode states $\rho^{T_B}$ is necessary and sufficient for their separability
\cite{peres,horodecki,simon}. The amount of entanglement can be measured through the
logarithmic negativity, defined as $E_{{\cal N}}=\log_{2}||\rho^{T_B}||$, being $||\rho^{T_B}||$
the modulus of the sum of the negative eigenvalues of $\rho^{T_B}$ \cite{vidal}. For any
two-mode Gaussian state, the  logarithmic negativity is $E_{{\cal
N}}=\max[0,-\log_{2}2\lambda_{-}]$, where $\lambda_{-}$ is the smallest symplectic
eigenvalue of $\rho^{T_B}$.

Because the system is connected to heat baths, the dissipating degree of freedom will
ultimately attain the thermal state corresponding to the system Hamiltonian. The
properties of entanglement in the thermal state are well studied, whence it is known that
only for very low temperatures entanglement is present between the two oscillators. 
Therefore we focus our attention on the time evolution of the decoherence process itself,
namely how fast an initially entangled state decays into a separable one, for separate
(Sect. \ref{decayseparate}) and  common (Sect. \ref{asympt}) baths.

\subsection{Entanglement decay time} 
\label{decayseparate}

We consider as a non-separable initial state the two-mode squeezed vacuum (\ref{TMS}),
with squeezing
parameter $r$. This state has an amount of entanglement proportional to $r$ and consists in
the orthogonal modes $x_\pm=(x_1\pm x_2)/\sqrt{2}$ being simultaneously squeezed and
stretched, respectively, by an amount $r$. Such a state, for infinite squeezing, gives the
maximally entangled state for continuous variables, long known due to the famous paper by
Einstein, Rosen and Podolsky \cite{EPR}.

\begin{figure}[h]
 \includegraphics[width=8cm]{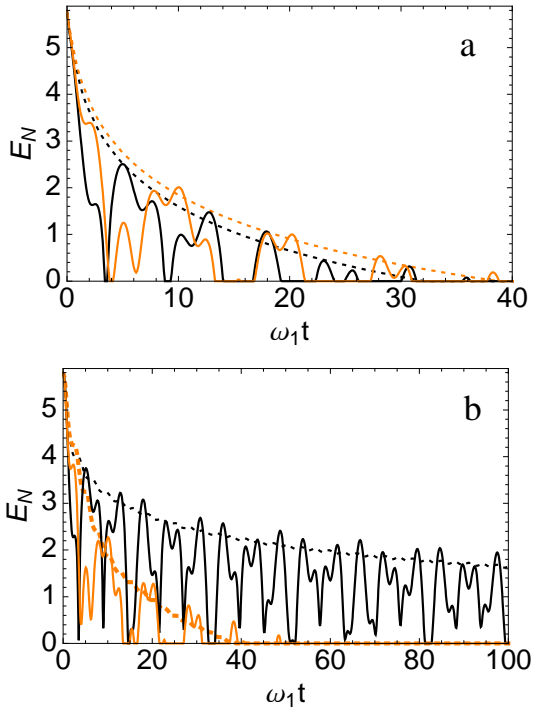}
\caption{(Color online). Time evolution of an entangled two-mode squeezed initial state
(\ref{TMS})  with $r=2$, in the
case of a) separate baths and b) common  bath. The parameters are $k_BT=10\omega_1$,
$\gamma=0.001\omega_1$, and cutoff frequency $\Lambda=50\omega_1$. The four curves
correspond to the points in figures \ref{fig0b} and \ref{fig1b}, namely A(dotted black),
B(solid black),C(dotted orange) and D(solid orange). Two features are apparent: in presence
of a nonzero coupling $\lambda$ (continuous lines)
entanglement oscillates strongly, and second, in the case of
separate baths any initial entanglement vanishes fast, while in the case of common bath it
can be made to survive by having identical frequencies.}
\label{figplots}
\end{figure}

\begin{figure}[h]
 \includegraphics[width=8cm]{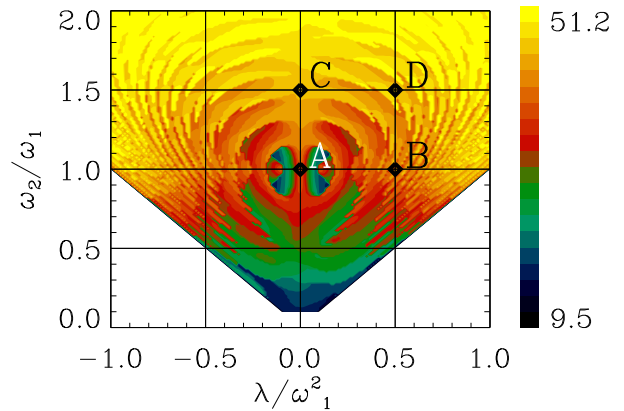}
\caption{(Color online). Parameter scan of $t_F$ (see text) as a function of the oscillators coupling and frequency detuning, in the case of separate baths. We have restricted the scan within a limited detuning ($\omega_2\leq2\omega_1$) as a result of the absence of any important features outside this region. In addition, the condition $\lambda<\omega_1\omega_2$ ensures reality of the eigenfrequencies in the problem. The behavior of $t_F$ is basically monotonic in $\omega_2/\omega_1$, with a superimposed arena-like shape, coming from the oscillatory nature of entanglement. See text for more details. The dots A, B, C, D are given in figure \ref{figplots} as examples.}
\label{fig0b}
\end{figure}

\begin{figure}[h]
 \includegraphics[width=9cm]{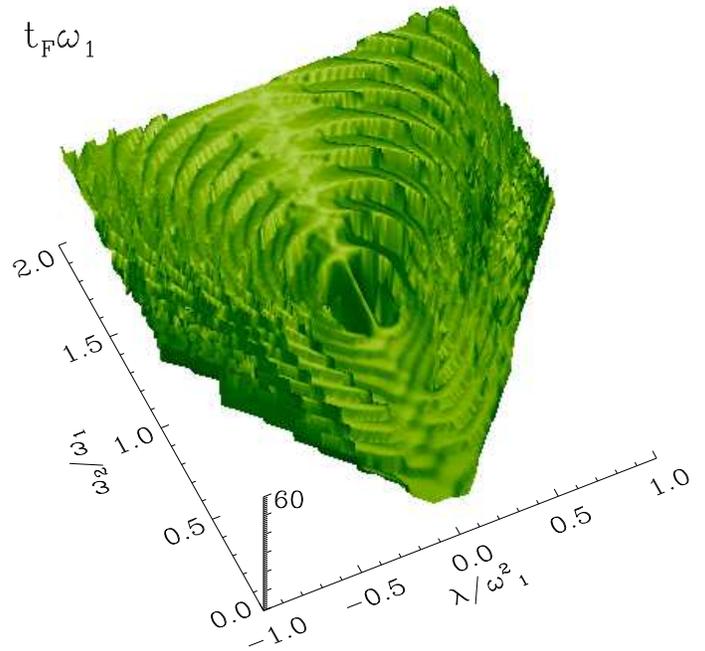}
\caption{(Color online). Three dimensional representation of figure \ref{fig0b}. Here the arena-like shape can be better appreciated. }
\label{fig0}
\end{figure}

The entanglement evolution in presence of separate baths is shown in 
Fig. \ref{figplots}a for different parameters choices, showing the effects of increasing
the coupling strength and the detuning. Entanglement, in general, decays with an
oscillatory behavior with oscillation amplitude depending on the $\lambda$ strength
\cite{galve}. We study the last time at which entanglement vanishes, $t_F$.
In Fig. \ref{fig0b} and \ref{fig0} we scan how long it takes for such an
entangled state (with $r=2$) to become separable when the oscillators detuning and
coupling strength are varied.  Indeed, the last time $t_F$ is represented in Fig. \ref{fig0}  for
different values of the ratio between frequencies ($\omega_2/\omega_1$) and coupling
between oscillators ($\lambda/\omega_1^2$).

For this choice of parameters, Fig. \ref{fig0b} shows that entanglement is present for
few tens of periods. Due to the scaling of  temperature, time, and couplings  with the
frequency of the first oscillator, $\omega_1$, it is clear that results shown in Figs.
\ref{fig0b} and \ref{fig0} are not symmetric by exchange of the oscillators role. In other
words, even if it is physically equivalent to have the first oscillator with half the
frequency of the second one or the second half the first one, due to the scaling, these
figures are not symmetric with respect to the line $\omega_2/\omega_1= 1$. 

Figure \ref{fig0b} shows two main features: (i) an increase of the survival time  $t_F$
when $\omega_2/\omega_1$ is increased; (ii)  an increase of $t_F$ with $|\lambda|$ in
absence of diversity ($\omega_2=\omega_1$), while for $\omega_2\gtrsim2\omega_1$ this
dependence  is weakened.  This can also be appreciated in the perspective presented in Fig.
\ref{fig0}.
Both features can be explained in terms of the eigenmodes of the system $Q_\pm$ 
(\ref{eigenmodes}) and their eigenfrequencies $\Omega_\pm$ (\ref{eigenfr}). Since the
eigenmodes do not interact with each other, they can be regarded as independent channels
for decoherence. The eigenmodes will only interact with near-resonant frequencies in the
bath. In addition, all variances at the thermal states they approach are
dependent upon the fraction $\Omega_\pm /2k_BT$. 

As far as it concerns the feature (i), in absence of 
 coupling ($\lambda=0$, implying $\Omega_{+,-}=\omega_{2,1}$) and for $\omega_2\to\infty$, 
 the `effective temperature' of the final thermal state reached by the eigenmode
 $Q_+$ will vanish, 
 $T_{\text{eff.},+}=k_BT/\omega_2\to 0$. 

The eigenmodes will therefore reach respectively
a thermal state ($T_{\text{eff.},-}=k_BT/\omega_1$) and a ground state ($T_{\text{eff.},+}=0$). 
Since the presence of entanglement is a competition between the reduced purities  of
individual oscillators and the total purity \cite{adesso}, just by improving
the final purity of oscillator $Q_+$ its time evolution has an overall higher purity,
thus making entanglement higher (and its survival time longer). This effect is equivalent
to reduce the real temperatures of the baths (while keeping everything else constant).
While the decoherence is given by the coupling $\gamma$ and is hence not reduced, the
reached final state is purer and thus entanglement is seen to survive longer.

As far as (ii) is concerned, for $\omega_2=\omega_1$ the eigenfrequencies of the system are
$\Omega_\pm=\sqrt{\omega_1^2\pm |\lambda|}$. By increasing $|\lambda|$ up to $\omega_1^2$
the eigenfrequency $\Omega_- $ vanishes and the amount of bath modes which have a similar
frequency, given by the spectral density, also vanishes ($J(\Omega_-\to 0)\to 0$). That is,
the amount of bath modes with which this degree of freedom interacts tends to zero, and
therefore the amount of decoherence suffered. This effect, however, is partially compensated
by the fact that a vanishing $\Omega_-$ would imply that this mode will reach a thermal
state of effective infinite temperature. Were it not so, the entanglement would survive
asymptotically, as in the case of common bath, where for $\omega_2=\omega_1$ the mode
$\Omega_-$ does not decohere.

We then see that all major effects related to entanglement decay between coupled
oscillators in presence of heat baths can be explained in terms of: 1) eigenfrequencies
(where they lie within the spectral density of the heat baths), and 2) effective
temperatures reached by the eigenmodes after thermalization.

A minor feature is that the increase of $t_F$ with $\lambda$ and
$\omega_2$ is not completely smooth due to entanglement oscillations and this gives rise to
the `arena'-shaped dependence that is clearer in Fig. \ref{fig0}. In general, the dependence of the
decoherence time on the oscillation frequency of the system is negligible for optics
experiments at ambient temperatures ($T_{\text{eff.},\pm}=k_BT/\Omega_\pm\simeq 0$), but becomes extremely important in presence of the
lower frequencies of mechanical oscillators \cite{giorgi2009}. 

Notice that we have restricted the coupling within the boundary
$|\lambda|<\omega_1\omega_2$ (triangle in the lower part of Fig. \ref{fig0b}), otherwise
one of the eigenfrequencies ($\Omega_-$) would become imaginary, meaning that trajectories
would be unbounded which, though not unphysical in principle, leads to leaks in any kind
of experiment. 

\subsection{Common bath}
\label{asympt}

 Let us compare now the previous results with the oscillators evolution with dissipation
through a common bath, Eq. (\ref{me_comm}).  This model can be considered a limit case
while separate baths would model the opposite one, when baths are completely
uncorrelated.   The presence of asymptotic entanglement in this situation has been studied
by \cite{asympt_ent_osc,paz-roncaglia,liu}, showing that entanglement can survive for
sufficiently low temperatures (or conversely for high enough initial squeezings).  In the
case with different frequencies Paz and Roncaglia noticed in Ref. \cite{paz-roncaglia2}
that this asymptotic behavior disappears as the detuning is increased, meaning that 
it is highly dependent on the frequency matching between oscillators and the fact that
each oscillator's bath can be regarded as `perfectly correlated' to the other one. In
physical realizations these two conditions will hardly be met and it is interesting to
know the effect of deviations from it. We have then studied the effects of frequency
diversity in the case of a common bath, looking again at the robustness of entanglement
in terms of the decay time $t_F$ (Fig.\ref{figplots}b). In Fig. \ref{fig1} we show results equivalent to what
represented in Fig. \ref{fig0} but for common bath. There we see that in the resonant case
entanglement never vanishes, and so the survival time is in that case infinite.  In Fig. 
\ref{fig1} diverging times $t_F$ are recognized along the line $\omega_1=\omega_2$.

When the oscillators frequencies begin to differ, survival diminishes fast to values
similar to the uncorrelated baths case. This phenomenon can be understood from the
master equation  (\ref{me_comm}). The fact that the coupling to the bath is
$\gamma(x_1+x_2)$ establishes that the mode
$x_-$ remains decoupled at all times, thus keeping its coherence
and contributing positively to a nonzero entanglement. If the frequencies are
different, that mode gets coupled to $x_+$ which is itself
coupled to the bath, and its decoherence is transferred to $x_-$. This way, both
modes are decohered and end up in a thermal state, with no presence of
entanglement. Hence the survival time is finite. Because of this, it is worthy to
stress the fact that even if the frequency difference is infinitesimal, decoherence
will eventually show up, even though it does so in a time inversely proportional to
the infitesimum. Thus, increasing the difference in frequencies just exacerbates
the transfer velocity of decoherence from mode $x_+$ to mode $x_-$. This argument
evidences how artificial the `equal frequencies' assumption is.

\begin{figure}
\includegraphics[width=8cm]{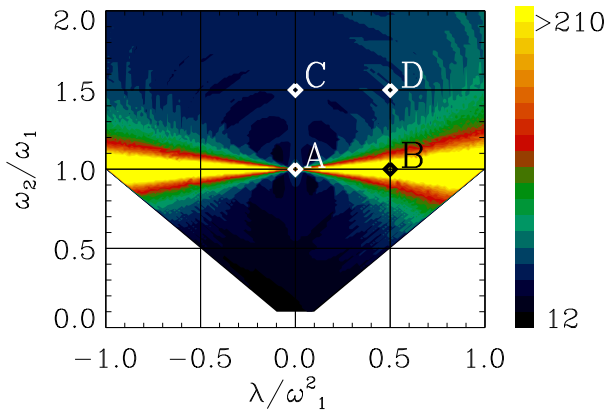}
\caption{(Color online). Parameter scan of $t_F$ (see text) as a function of the oscillators coupling and frequency detuning, in the case of common bath. As seen here, $t_F$ is huge whenever the detuning is small, and in the order of tens of periods for high detuning. As expected, at resonance one of the eigenmodes decouples from the baths, leading to an infinite $t_F$. We have truncated the plot at heights of $\omega_1 t_F=210$, otherwise this `mountain riff' is infinitely high (but only strictly infinite when $\omega_1=\omega_2$).}
\label{fig1b}
\end{figure}

\begin{figure}
\includegraphics[width=9cm]{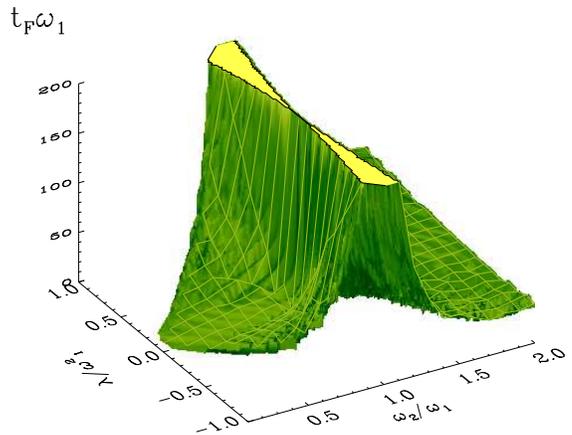}
\caption{(Color online). 3D representation of Figure \ref{fig1b}. Apart from the `riff' at equal frequencies, the shape is very similar to the case of separate baths, with an arena-like bowl shape. We have drawn a mesh to guide the reader's eye. Again, heights over $t_F=210/\omega_1$ have been truncated (yellow region).}
\label{fig1}
\end{figure}

We point out that asymptotic entanglement can be found also in
presence of frequency diversity and is not related to the symmetry present for
$\omega_1=\omega_2$. As a matter of fact, if the couplings to the common bath are
different for each oscillator, i.e. $\gamma_1\neq\gamma_2$, we can restore
asymptotic entanglement even off-resonance, for $\omega_1\neq\omega_2$. 
By noticing the relation between $x_{1,2}$ with
the eigenmodes of the system $Q_\pm$, we see that the mode $Q_+$ can be uncoupled
from the bath when the angle $\theta$ fulfills
\begin{equation}\label{cond_theta}
\cos{\theta}=\frac{\gamma_1}{\sqrt{\gamma_1^2+\gamma_2^2}}
\end{equation}
with $\theta=\theta(\omega_1,\omega_2,\lambda)$ given in the Appendix.
Thus, half of the system is kept in a pure state, and we can show that entanglement
survives asymptotically exactly in the same fashion as it did for equal frequencies
above. In other words, with different couplings to the baths together
with different frequencies, the system can retain asymptotic entanglement if Eq.
(\ref{cond_theta}) is fulfilled. In any case this phenomenon is as unique as the
`equal frequencies' one, and requires fine-tuned parameters (tuning bath couplings or
frequency difference) in order to be observable. It should then be regarded as
exceptional too.  Our argument clarifies that asymptotic entanglement
is $not$ a consequence of a symmetry in the system, but   {\it comes from finely
tuned decoherent-free degree of freedom
of the system.}

\subsection{Twin oscillators}

\begin{figure}[h]
\includegraphics[width=8cm]{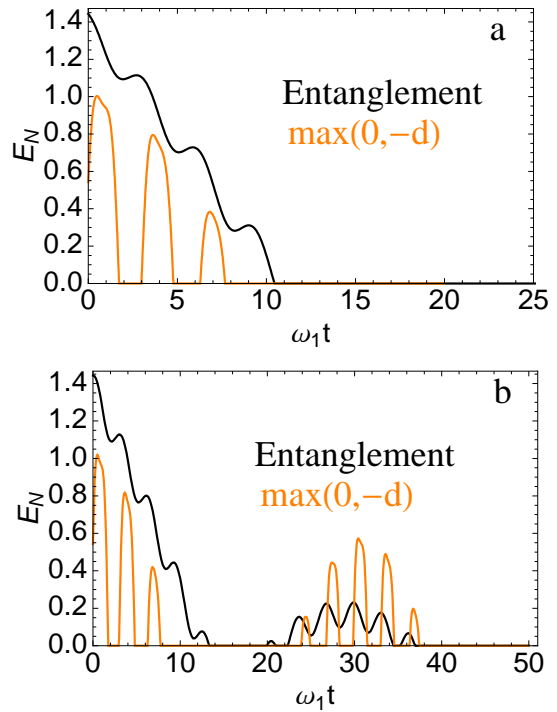}
\caption{(Color online). Comparison between the time evolutions of entanglement (black) and the correlation measure 
$\rm{max}(0,-\langle:(n_1-n_2)^2:\rangle$) (orange), for the case of a) separate 
baths and b) common bath. We have used equal frequencies, coupling $\lambda=0.1\omega_1^2$, 
damping $\gamma=0.001\omega_1$, temperature $k_BT/\omega_1=10$ and initial squeezing parameter 
$r=0.5$. The envelope of the correlation measure seems to be closely related to the entanglement, 
except for a delay of half a period.}
\label{fig2}
\end{figure}

We have considered entanglement as an indicator of the quantumness of our system, measurable for the
family of (Gaussian) states considered here. The discrimination between the predictions of classical
and quantum theories for coupled harmonic oscillators has been largely studied also in optics. In that
context there have been many theoretical predictions experimentally confirmed, focusing on the violation
of different classical inequalities, or the positivity of variances
\cite{ineq}. An example is the variance of the difference of the occupation numbers:
\begin{equation}
d=\langle:(n_1-n_2)^2:\rangle\label{vard}
\end{equation}
where, as usual, $n_i$ is the occupation number operator of each oscillator.  $d$ has been considered
in two mode squeezed state generated by parametric oscillators in optics, to characterize twin beams
\cite{loudon}.
The quantum character of the correlations in the occupation numbers comes from the negativity of the
variance $d$ and can only follow from the negativity of the corresponding quasi-probability, the
Glauber-Sudarshan representation in this case.  In analogy with the optical case, we consider here
$twin$ oscillators looking at the temporal dynamics of $d$. 
The correspondent fourth order moments can be obtained from the covariance
matrix, because the states we are dealing with are Gaussian. In Fig. \ref{fig2} we plot two examples
comparing the evolution of this variance and entanglement for the common/separate baths cases. We have
taken only the negative part of this correlation -identifying quantum behavior- and inverted it
(plotting $\max(0,-d)$ ) for ease of comparison with entanglement. For the initial entangled
state (\ref{TMS}),
$d=-2\mu/(1-\mu)$, always negative. In other words, for squeezed states the occupation number of one 
oscillator determines the other one.
We observe that starting from this value $d$ decays with large oscillations  
and that the sudden deaths of
entanglement and of this correlation coincide up to a fraction of a period. Still entanglement
evolution is smoother and twin oscillators temporarily loose their quantum correlations even for 
entangled oscillators. 

\section{Non-Markovian evolution}
\label{non-Markovian}

Up to now we have considered the Markovian master equation, in the sense that the
time-dependent coefficients related to the heat bath have been replaced by their
asymptotic value, obtained  by integrating up to infinite time. This implies a
complete absence of memory in the bath, which does not retain
instantaneous information on the dynamics of the two oscillators. However, this
simplification can be dropped, and we are left with a complete description {\it
within the weak coupling approximation}.  In this
section we analyze how these corrections affect the entanglement evolution we
discussed so far. 

 In Ref. \cite{maniscalco}, a comparison has been made considering temperatures 
two orders of magnitude greater than the frequency cut-off. The authors found that, 
when the frequency of the oscillators falls inside the bath spectral density, entanglement
 persists for a longer time than in a Markovian channel. When there is no resonance 
between reservoir and oscillators, non-Markovian correlations accelerate
 decoherence and generate entanglement oscillations.
 
\begin{figure}[h]
\includegraphics[width=7cm]{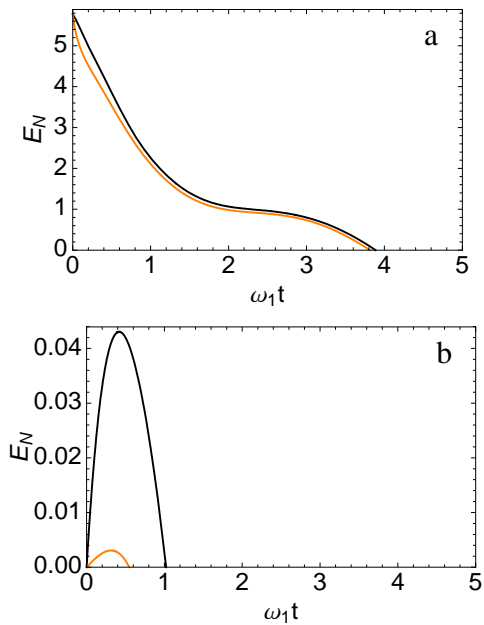}
\caption{(Color online). Comparison between Markovian (orange) and non-Markovian (black) entanglement dynamics.
 The system parameters are $\omega_1=\omega_2$, $\lambda=0.2\omega_1^2$,
 $\gamma=0.01\times 2/\pi\omega_1$,  $\Lambda=20\omega_1$, and  $k_BT=10\omega_1$. The initial squeezings are a) $r=2$ and b) $r=0$.}
\label{nmars}
\end{figure}

As it is known, when all the important frequencies
are much lower than the cut-off, the density of states of Eq.
(\ref{JOhm}) is expected to generate almost memoryless friction  \cite{weiss}. Then,  
taking $ \omega_1\ll\Lambda$, we compared Markovian and non-Markovian entanglement dynamics
for different choices of the parameters of the system and the bath ($\omega_2/\omega_1$,
$\lambda/\omega_1^2$, $\gamma/\omega_1$, $k_BT/\omega_1$, $r$). The following considerations
are valid for separate baths as well as for a common environment. In the case of very small coupling
($\gamma\sim 10^{-3}\omega_1$) non-Markovian corrections are practically unobservable
independently on the value of the system parameters up to relevant temperatures. To observe
some appreciable deviations, we need to reach temperatures of the order of $10\omega_1$. This
result is understood considering that for small values of the system-bath coupling the role
played by the bath itself is highly reduced. Since non-Markovian corrections to the values of
the coefficients are relevant only during the first stage of the evolution, it can be also
predicted that if the initial state is robust enough (as, for instance, in the case of
squeezing $r=2$ discussed in the previous sections), non-Markovianity  will have marginal
effects. It is in fact true that, in the case of $\gamma \sim 10^{-3}\omega_1$ and  $T\sim 10
\omega_1$, a correction to the entanglement decay time can only be observed  assuming low
initial squeezing ($r\lesssim0.1$). When $\gamma$ starts to increase, it is possible to deal with
cases where some observable corrections due to non-Markovianity emerge also in the
low-temperature regime. In particular, the case of equal frequencies turns out to be the more
sensitive. To give an example, in Fig. \ref{nmars}  we compared the two evolutions for two
different values of $r$. Starting from $r=2$, non-Markovian corrections are almost
negligible, giving corrections of few percents to the value of $t_F$
found within the Markovian approximation. 
In order to observe a noticeable difference, we must start with a factorized
state ($r=0$). As it can be observed in Fig.  \ref{nmars}, in this case, while memoryless
baths are able to support a very small quantity of entanglement, the initial kick given by
non-Markovian coefficients enhances entanglement generation, its maximum amount is amplified
of about one order of magnitude, and also its death time is increased. It is however worth
noting that the absolute value reached is, in any case, very low. 

\section{Discussion and conclusion}
\label{concl}

During last years, a series of papers has investigated entanglement dynamics of  coupled
harmonic oscillators in contact with thermal environments. Nevertheless, a comprehensive
analysis around the role played by the diversity of the two oscillators was missing.

We started our discussion with the case of separate baths, which seems to be more
physically meaningful.  As one should expect, after a transient, unless the temperature is
very low, the initial two-mode squeezed state  $| \Psi_{TMS}\rangle$ becomes separable.
Studying the decay time, we observed a series of interesting features that we have
qualitatively explained. First of all, by increasing $\omega_2/\omega_1$ and by keeping
unchanged all other parameters, entanglement survival time increases as well. In the limit
of $\omega_2/\omega_1\gg1$, $\Omega_+$ and $\Omega_-$ tend, respectively, to $\omega_2$
and $\omega_1$. Since the asymptotic thermal state the system is reaching depends itself on
the frequencies, the purity of the final state is enhanced together with the coherence
time necessary to reach it. On the other hand, near $\omega_2=\omega_1$,  entanglement
survival time can be raised also by increasing $|\lambda| $. When $|\lambda|\rightarrow
\omega_1^2$, $\Omega_-$ is close to zero. Here, we argue that the behavior we observe is the result
of two competing effects. If, on one side, the thermal equilibrium state  is less pure
because of the presence of a vanishing frequency, on the other side, the number of
decohering channels goes to zero, since $\Omega_-$ falls in a poorly populated region of
the spectral density, increasing in this way the time necessary to reach the thermal
equilibrium.

If the two oscillators share a common bath, it becomes possible the observation of
asymptotic entanglement at relevant temperatures. One of the two eigenmodes of the system
can result in fact decoupled from the environment.  As a consequence of this decoupling,
the asymptotic density matrix will result as the direct sum of the matrix of pure state
and of a thermal state, corresponding to the mode interacting with the bath. When the two
oscillators have identical frequencies and bath couplings, the
mode $x_-$ gets decoupled independently of the value of $\lambda$, giving rise to
the plateau we observe in Figs. \ref{fig1b}, \ref{fig1}. It is important to stress that
this regime can be achieved not only when the oscillators are resonant, but also in
presence of frequency detuning, through a proper engineering of the system-bath
interaction. Conversely, if their bath couplings are different
$\gamma_1\neq\gamma_2$, frequencies $\omega_1$ and $\omega_2$ can be found so that this
special regime is achieved.

We also studied the dynamical evolution of the degree of quantumness of the system by
means of the quantity $d$, introduced in Eq. (\ref{vard}).  The negativity of this
quantity is a direct evidence of the negativity of the Glauber-Sudarshan quasi-probability
distribution of the state. The agreement between the behavior of $d$ and the entanglement
is worth to be investigated, since a tight relation between them is not known. 

As for the Markovian versus non-Markovian debate, we compared the two behaviors in a wide
range of parameters. While in Ref. \cite{maniscalco} the authors investigated what changes
when the frequency of the oscillators goes outside  from the spectral distribution of the
bath in the high temperature regime, we tried to do an extensive analysis  for
$\omega_1,\omega_2\ll\Lambda$. Our results indicate that non-Markovian corrections can be
observed, and are important,  only for a reduced subset of initial conditions. To see
them, we must put a relevant coupling between oscillators and bath, $\omega_1$ should be
close to $\omega_2$, and the initial state should be (almost) disentangled. Since the
master equation  has been obtained in the weak-coupling limit, the first of this
assumptions seems at least questionable. If we release only one of them, Markovian and
non-Markovian evolutions are almost identical.


\acknowledgments{
We acknowledge funding from Juan de la Cierva program and from CoQuSys project.}

\begin{appendix}

\section{Derivation of master equation}
A system-bath model is usually described through the Hamiltonian $H=H_S + H_B + V$, where $H_S$ refers to the system and $H_B$ to the bath, and where $V$ is the term containing the interaction. Generally, the coupling term can be written as  $V=\sum_k S_k\otimes B_k$, where  $ S_k$ denote system operators and $ B_k$ bath operators.
The master equation for the reduced density matrix to the second order in the
 coupling strength (for a general discussion see for instance \cite{gardiner}) is given by
\begin{eqnarray}
 \frac{d \rho}{dt}&=&-i[H_S,{\rho}]\nonumber
\\&-&\int_0^td\tau\sum_{l,m}\big\{ C_{l,m}(\tau)[S_l \tilde{S}_m(-\tau)\rho-\tilde{S}_m(-\tau)\rho S_l] \nonumber
\\&+& C_{m,l}(-\tau)[\rho\tilde{S}_m(-\tau)S_l -S_l\rho\tilde{S}_m(-\tau)]\big\}.\label{a1}
\end{eqnarray}
Here, $C_{l,m}(\tau)$ are the bath's correlation functions, defined by
\begin{equation}
 C_{l,m}(\tau)={\rm Tr}_B [\tilde{B}_l(\tau)\tilde{B}_m(0)R_0],\label{a2}
\end{equation} 
being $R_0=\left( e^{-\beta H_B}/{\rm Tr e^{-\beta H_B}}\right) $ the equilibrium density matrix of the bath. In Eqs. (\ref{a1},\ref{a2})
 the tilde indicates the interaction picture taken with respect to $H_0=H_S+H_B$. For instance, $\tilde{B}_l(\tau)=e^{i H_0 \tau} B_le^{-i H_0 \tau}$. 

In the following, the correlation functions will appear combined as $C_{l,m}^A(\tau)= C_{l,m}(\tau)+ C_{m,l}(-\tau)$ and $C_{l,m}^C(\tau)= C_{l,m}(\tau)- C_{m,l}(-\tau)$.

\subsection{Separate baths}
In our model, whose Hamiltonian is $H^{sep}$, we can identify $B_1=\sum_{k}\lambda_{k}^{(1)}X_{k}^{(1)}$ and $B_2=\sum_{k}\lambda_{k}^{(2)}X_{k}^{(2)}$ and, correspondingly, $S_1=x_1,S_2=x_2$ . Their expressions in the interaction picture are
\begin{equation}
 \tilde{B}_i=\sum_{k}\lambda_{k}^{(i)}\left[ X_{k}^{(i)}\cos \Omega_{k}^{(i)}t+P_{k}^{(i)}\dfrac{ \sin \Omega_{k}^{(i)}t}{m \Omega_{k}^{(i)}}\right] 
\end{equation}
and $\tilde{x}_i(\tau)=\alpha_{i1}x_1+\alpha_{i2}x_2+\beta_{i1}p_1+\beta_{i2}p_2$, with
\begin{eqnarray}
 \alpha_{11}&=&\cos^2\theta \cos\Omega_- \tau+\sin^2\theta \cos\Omega_+ \tau,\label{int1}\\
 \alpha_{12}&=&\frac{\sin2\theta}{2}(\cos\Omega_+ \tau-\cos\Omega_-  \tau),\label{int2}\\
 \beta_{11}&=&\frac{\cos^2\theta }{\Omega_-}  \sin\Omega_- \tau+\frac{\sin^2\theta }{\Omega_+}  \sin\Omega_+ \tau,\label{int3}\\
 \beta_{12}&=& \frac{\sin2\theta}{2}( \frac{\sin\Omega_+ \tau}{\Omega_+}  -\frac{ \sin\Omega_- \tau}{\Omega_-}  )\label{int4},\end{eqnarray}
and $\alpha_{22}(\theta)=\alpha_{11}(\pi/2-\theta)$, $\beta_{22}(\theta)=\beta_{11}(\pi/2-\theta)$, $\alpha_{21}=\alpha_{12}$, and $ \beta_{21}=\beta_{12}$.
The coefficients appearing in Eqs. (\ref{int1}-\ref{int4}) are defined as
\begin{equation}
 \theta=\frac{1}{2}\arctan \left( \frac{2\lambda}{\omega_2^2-\omega_1^2}\right) 
\end{equation}
and
\begin{equation}\label{eigenfr}
 \Omega_\pm=\sqrt{\frac{\omega_1^2+\omega_2^2}{2}\pm\frac{\sqrt{4\lambda^{2}+ 
 (\omega_2^2-\omega_1^2)^2}}{2}}.
\end{equation}
The normal modes for the position are
\begin{eqnarray}\label{eigenmodes}
Q_{+}=\cos\theta x_2+\sin\theta x_1,\\ Q_{-}=\cos\theta x_1-\sin\theta x_2,
\end{eqnarray}
while for the momentum
\begin{eqnarray}
P_{+}=\cos\theta p_2+\sin\theta p_1,\\ P_{-}=\cos\theta p_1-\sin\theta p_2.
\end{eqnarray}
Notice that, for $\omega_{1}=\omega_{2}$, $\theta=\pi/4$. This implies $ \alpha_{11}= \alpha_{22}$ and $ \beta_{11}= \beta_{22}$.

Since the two baths are identical and uncorrelated, we have $C_{1,1}(\tau)=C_{2,2}(\tau)=C(\tau)$ and $C_{1,2}(\tau)=C_{2,1}(\tau)=0$. From the knowledge of the density of states $J(\Omega)$ it is possible to obtain the explicit expression of $ C^C(\tau)$ and $ C^A(\tau)$:
\begin{eqnarray}
  C^C(\tau)&=&-i\int_0^\infty d\Omega J(\Omega) \sin \Omega \tau, \\ C^A(\tau)&=&\int_0^\infty d\Omega J(\Omega) \cos \Omega \tau \coth\frac{\Omega}{2 k_B T}.
\end{eqnarray}
 The master equation reads
\begin{eqnarray}
\frac{d \rho}{dt}&=&-i[H_S,{\rho}]-\int_0^t d\tau\sum_{i=1,2} \big\{[x_i,\{x_i(-\tau),\rho\}]\frac{C^C(\tau)}{2}\nonumber\\&+&[x_i,[x_i(-\tau),\rho]]\frac{C^A(\tau)}{2}\big\}.
\end{eqnarray}
By using expressions (\ref{int1}-\ref{int4}), it can be written as in Eq. (\ref{me_sep}),
with
\begin{eqnarray}
\epsilon_{ij}^2&=&-i\int_0^t d\tau \alpha_{ij}C^C(\tau),\label{coe}\\
 D_{ij}&=&\int_0^t  d\tau \alpha_{ij}C^A(\tau),\label{cod}\\
 F_{ij}&=&\int_0^t  d\tau \beta_{ij}C^A(\tau),\label{cof}\\
\Gamma_{ij}&=&i\int_0^t  d\tau \beta_{ij}C^C(\tau).\label{cog}
\end{eqnarray}
The considerations about symmetry given in Sec. \ref{model} can be understood in terms of the explicit form of the master equation. The substitution $\lambda\rightarrow-\lambda$ is equivalent to sending $\theta$ to $-\theta$. Then, also $ \alpha_{12}$ and $\beta_{12}$ change their sign. But, since these coefficients multiply one operator (position or momentum) of the first oscillator and one operator of the second one, the canonical transformation $U$ compensates the change of sign, and the evolution of $\rho$ is unchanged.

The explicit form of $\epsilon_{ij},D_{ij},\Gamma_{ij},F_{ij}$, in the Markovian case, can be calculated using the equalities
\begin{equation}
 \lim_{t\rightarrow\infty}\int_0^t d\tau e^{-i(\omega-\omega_0)\tau}=\pi \delta (\omega-\omega_0)+i\frac{P}{\omega-\omega_0},
\end{equation}
where $P$ denotes the Cauchy principal value, and
\begin{equation}
\coth{\frac{\omega}{2 k_B T}}=2 k_B T\sum_{n=-\infty}^{+\infty}\frac{\omega}{\omega^2+\nu_n^2},
\end{equation}
with $\nu_n=2\pi n k_BT$.

From the master equation, a set of closed equations of motion for the average values of the second moments can be derived ($i,j=1,2$):
\begin{widetext}
\begin{eqnarray}
  \frac{d \langle x_{i}x_{j} \rangle}{dt}&=&\langle p_{i} x_{j}+p_{j} x_{i}\rangle\\
\frac{d \langle p_{i}p_{j} \rangle}{dt}&=&-\frac{1}{2}(\omega_{i}^{2}+\epsilon_{ii}^2)\langle \{x_{i}, p_{j}\}\rangle-\frac{1}{2}(\omega_{j}^{2}+\epsilon_{jj}^2)\langle \{ x_{j}, p_{i}\}\rangle\nonumber\\
&-&\frac{1}{2}(\lambda+\epsilon_{12}^2)\sum_{k=1}^2[\langle \{x_{k},p_{i}\} \rangle(1-\delta_{k,j})+ \langle \{x_{k},p_{j}\}\rangle(1-\delta_{k,i})]\nonumber\\
&-&(\Gamma_{ii}+\Gamma_{jj})\langle p_{i}p_{j} \rangle-\Gamma_{12}\sum_{k=1}^2[\langle p_{k}p_{i} \rangle(1-\delta_{k,j})+ \langle p_{k}p_{j}\rangle(1-\delta_{k,i})]+D_{ij}\\
 \frac{d \langle \{x_{i},p_{j}\} \rangle}{dt}&=&2\langle p_{i} p_{j}\rangle-2(\omega_{j}^{2}+\epsilon_{jj}^2)\langle x_{i} x_{j}\rangle-2(\lambda+\epsilon_{12}^2)\left[ \langle x_{i}^{2}\rangle(1-\delta_{ij})+\delta_{ij}\langle   x_1 x_2\rangle\right] \nonumber\\
&-&\Gamma_{jj}  \langle \{x_{i},p_{j}\} \rangle-\Gamma_{12}\langle \{x_{i},p_{i}\} \rangle+F_{ij}
\end{eqnarray}

\end{widetext}

\subsection{Common bath}

In this case, $C_{1,1}(\tau)=C_{2,2}(\tau)=C_{1,2}(\tau)=C_{2,1}(\tau)$. Then, the master
equation can be written as has
\begin{eqnarray}
\frac{d \rho}{dt}&=&-i[H_S,{\rho}]-\int_0^t d\tau\sum_{i,j=1,2} \big\{[x_i,\{x_j(-\tau),\rho\}]\frac{C^C(\tau)}{2}\nonumber\\&+&[x_i,[x_j(-\tau),\rho]]\frac{C^A(\tau)}{2}\big\}.
\end{eqnarray}
and, using Eqs. (\ref{coe}-\ref{cog}), assumes the form given in Eq. (\ref{me_comm})
with $\bar D_{ii}=D_{ii}+D_{12}$  and similarly for $\epsilon^{2},\Gamma,F$. The equations of motion for the second moments read ($ i,j=1,2$)
\begin{widetext}
\begin{eqnarray}
 \frac{d \langle x_{i}x_{j} \rangle}{dt}&=&\langle p_{i} x_{j}+p_{j} x_{i}\rangle\\
\frac{d \langle p_{i}p_{j} \rangle}{dt}&=&-\sum_{k=1}^2 (\omega_{k}^{2}+\bar{\epsilon}_{kk}^2)( \langle x_{k}p_{j} \rangle\delta_{ik}+ \langle p_{i}x_{k} \rangle\delta_{jk})\nonumber\\
&-&\frac{1}{2}\sum_{k=1}^2 (\lambda+\bar{\epsilon}_{ii}^2 )\left[  \langle \{x_{k},p_{i}\}\rangle(1-\delta_{kj})+
\langle \{x_{k},p_{j}\}\rangle(1-\delta_{ki}) \right] +\frac{1}{2}(\bar D_{ii}+\bar D_{jj})\nonumber\\
&-&(\bar{\Gamma}_{ii}+\bar{\Gamma}_{jj})\langle p_{i}p_{j} \rangle-\sum_{k=1}^2\bar{\Gamma}_{kk}[\langle p_{i}p_{k} \rangle(1-\delta_{kj})+\langle p_{j}p_{k} \rangle(1-\delta_{ki})]\\
 \frac{d \langle \{x_{i},p_{j}\} \rangle}{dt}&=&2\langle p_{1} p_{2}\rangle-2(\omega_{j}^{2}+\bar{\epsilon}_{jj}^2)\langle x_{i} x_{j}\rangle\nonumber\\
&-&2\sum_{k=1}^2(\lambda+\bar{\epsilon}_{kk}^2 ) \langle x_{i}x_k\rangle (1- \delta_{ik})\delta_{ij}
-2(\lambda+\bar{\epsilon}_{ii}^2 ) \langle x_{i}^2\rangle (1-\delta_{ij})\nonumber\\&-&\bar\Gamma_{jj} \langle\{ x_{i},p_{j}\} \rangle-\sum_{k=1}^2\bar\Gamma_{kk}\langle \{x_{i},p_{k}\} \rangle (1- \delta_{ik})+\bar F_{ii}\nonumber\\
\end{eqnarray}
\end{widetext}

\end{appendix}

\end{document}